\documentclass{cernyrep}
\usepackage[T1]{fontenc}
\usepackage{graphicx}
\usepackage{cancel}
\usepackage[bookmarks, colorlinks=true, linktoc=page, pdftex, linkcolor=black, citecolor=black, urlcolor=blue]{hyperref}

\usepackage{fancyhdr}
\fancyhfoffset{4 mm}
\fancypagestyle{ARTTITLE}{%
\fancyhf{} 
\lhead{Proceedings of the CERN--Accelerator--School course on
{\it Introduction to Accelerator Physics}}
\lfoot{Available online at \url{https://cas.web.cern.ch/previous-schools}}
\rfoot{\thepage\hspace*{3mm}}
 
}
\usepackage{amsmath}

\newcommand{\vect}[1]{\boldsymbol{#1}}

\usepackage{varwidth}
\usepackage{xcolor}

\renewcommand{\vec}[1]{\ensuremath{\mathbf{#1}}}

\newcommand{\myfrac}[2]{\ensuremath{\frac{#1}{#2}}}

\newcommand{\timediff}[1]{\ensuremath{\myfrac{\partial #1}{\partial t}}}

\newcommand{\nablabold}{\ensuremath{\mathbf{\nabla}}}

\newcommand{\divvec}[1]{\ensuremath{\nablabold \cdot #1}}

\newcommand{\curlvec}[1]{\ensuremath{\nablabold \times #1}}

\newcommand{\chargedensity}{\ensuremath{\rho}}

\newcommand{\electric}{\ensuremath{e}}

\newcommand{\electricfield}{\ensuremath{\vec{E}}}

\newcommand{\electricchargedensity}{\ensuremath{\chargedensity_{\electric}}}

\newcommand{\magneticfield}{\ensuremath{\vec{B}}}

\begin{document}

\title{Introduction to Electromagnetism}
\author{Dr. Irina Shreyber}
\institute{CERN, Geneva, Switzerland}

\begin{abstract}
The purpose of this course is to provide an introduction to Electromagnetic Theory. The foundations of electrodynamics starting from the nature of electrical force up to the level of Maxwell equations solutions are presented. It starts with the introduction of the concept of a field, which plays a very important role in the understanding of electricity and magnetism. In addition, moving electric charge is discussed as a topic of special importance in accelerator physics. 

\end{abstract}
\maketitle
\thispagestyle{ARTTITLE}

\section{Content of the course}
The topics that will be covered in this lecture are the following:
\begin{enumerate}
\setcounter{enumi}{1}
  \item Introduction
    \begin{itemize}
        \item Introduction to Fields
        \item Charge and Current
        \item Conservation Law
        \item Lorentz Force
        \item Maxwell's Equations
    \end{itemize}
  \item Electrostatics 
   \begin{itemize}
     \item Coulomb Force
     \item Electrostatic Potential
     \item Principle of Superposition
     \item Continuous distribution of charges
   \end{itemize}
  \item Magnetostatics
     \begin{itemize}
     \item Steady Current
     \item Ampère's Law
     \item Vector Potential
     \item Biot-Savart Law
     \item Motion of a charged particle
   \end{itemize}

  \item Electromagnetism
     \begin{itemize}
     \item Faraday's Law of Induction
     \item Wave Function
     \item Propagation of electromagnetic waves in a conductor
     \item Propagation of electromagnetic waves in a  highly conductive materials
  
   \end{itemize}
\end{enumerate}

All material covered in this lecture and details of the calculations involved can be found in classical textbooks for electromagnetic theory (see for example  \cite{bib:Griths, bib:Jackson, bib:Zangwill}. In writing this lecture I have benefited from lectures given in previous CERN schools~\cite{bib:CAS00-Lahanas, bib:CAS19-Latina}, as well as from the lectures on electromagnetism of Andy Wolski~\cite{bib:AE-Wolski}, David Tong~\cite{bib:DTong}, Robert de Mello Koch and Neil Turok~\cite{bib:Koch}.

\section{Introduction}
\subsection{Introduction to fields}

We start by looking at the gravitational force exerted by the earth on a particle, which allows to introduce the concept of a field.  From Newton it is known that if a force $\vect{F}$ (here and later the bold notation for vectors is used) acts on a particle of mass $m$, then the particle will experience an acceleration $\vect{g}$:
\[\vect {F} = m\vect{g} = m \frac{d^2\vect{x}}{dt^2}\]

By measuring the forces acting on the particle, we discover that at each different position in space the acceleration is different. This implies that the magnitude of the force changes at each different position in space at a different time. 

In physics, a \emph{field} is a dynamical quantity that has a value at each location in space and at each instant in time. This means that instead of saying that the earth exerts a force on a falling object, it is more useful to say that the earth sets up a gravitational force field. A \emph{force} in modern physics, means an intricate interplay between particles and fields.

To describe the force between charged particles an electric field is introduce by copying the description of the gravitational force field. As charged particles exert forces on each other, we write the same as for the gravitational field:
\[ \vect{F} = q\vect {E}\]

The charge $q$ replaces the mass $m$.  $q$ is a single number associated with the object that experiences the field. The electric field $\vect{E}$ is what replaces the gravitational field $\vect{g}$. So we have replaced a number with a number $m \rightarrow q$ and a field with a field ($\vect{g} \rightarrow \vect{E}$). We are splitting things up into a source that produces a field and an object that experiences the field.

To describe the force of electromagnetism, we need to introduce two fields, each of which is a three-dimensional vector. They are called the electric field $\vect{E}$, which is described above, and the magnetic field $\vect{B}$. 
\[ \vect{E}(\vect{x}, t) ;  \vect{B}(\vect{x}, t)\]

The charged particles create both electric and magnetic fields. The electric and magnetic fields guide the charged particles. This motion, in turn, changes the fields that the particles create. Roughly speaking, an electric field accelerates a particle in the direction $\vect{E}$, while a magnetic field causes a particle to move in circles in the plane perpendicular to $\vect{B}$.

\subsection {Charge and Current}

We measure the charge $q$ in \emph{Coulomb}, and charge can be positive or negative. In SI units the charge of a single proton is about $1.6\times 10^{-19}$C. 

Often in particle physics we simply count charge as $q = ne$ with $n \in \mathbb{Z}$. Then electrons have charge $-1$, while protons have charge $+1$ and neutrons have charge 0.

To move from the dynamics of point particles onto the dynamics of continuous objects known as fields, the charge density is introduced,
\[ \rho(\vect{x}, t) \]
defined as charge per unit volume. The total charge $Q$ in a given region $V$ is then simply
\[ Q = \int_V d^{3}x \rho(\vect{x}, t). \]

In most situations, smooth charge densities are considered, which can be thought of as arising from averaging over many point-like particles. 

Electric fields are produced by the static charges, magnetic fields is set up by currents, i.e. moving charges. To describe the movement of charge from one place to another a quantity known as \emph{the current density} is used $\vect{J}(\vect{x}, t)$. It is defined as follows: for every surface $S$, the integral
\[ I = \int_S \vect{J} \cdot d\vect{S} \]
counts the charge per unit time passing through surface $S$, ($d\vect{S}$ is the unit normal to $S$). The quantity $I$ is called the \emph{current}. In this sense, the current density is the current-per-unit-area.

\subsection{Conservation Law}

The most important property of electric charge is that it is conserved, i.e. the total charge in a system cannot change. 

The property of local conservation means that $\rho$ can change in time only if there is a compensating current flowing into or out of that region. This is expressed in the \emph{continuity equation}
\[   \frac{d\rho}{dt} + \vec{J} = 0 .\]

To see why the continuity equation captures the right physics, it is best to consider the change in the total charge $Q$ contained in some region $V$
\[  \frac{dQ}{dt} = \int_V{d^{3}x \frac{d\rho}{dt}} = - \int_V{d^{3}x \vec{J}} = - \int_S \vect{J} \cdot  d\vect{S} .\]
The minus sign is there to ensure that if the net flow of current is outwards, then the total charge decreases.

If there is no current flowing out of the region, then $\frac{dQ}{dt} = 0$. This is the statement of (global) \emph{conservation of charge}.  In many applications $V$ is taken to be all of space with both charges and currents localised in some compact region. This ensures that the total charge remains constant.

\subsection{Lorentz Force}

The position $\vect{r}(t)$ of a particle of charge $q$ is dictated by the electric and magnetic fields through the \emph{Lorentz force law}. The force acting on a charge $q$, which is at rest within an electric field, is 
\[ \vect{F} = q\vect{E}. \] 

Also the force acting on a small wire element $\vect{dl}$, carrying electric current $I$, which is placed in a magnetic field, is 
\[
\vect{F} = I \vect{dl} \vect{B}.
\]

These two suggest that for a charge $q$ moving with velocity $\dot{\vect{r}}$, the total force acting (Lorentz force) on it: 
\[ 
\vect{F} = q (\vect{E} + \dot{\vect{r}} \times \vect{B}) . 
\]

The Lorentz force law can be also written in terms of the charge distribution $\rho(\vect{x}, t)$ and the current density $\vect{J}(\vect{x}, t)$. In terms of the force density $\vect{f}(\vect{x}, t)$, which is the force acting on a small volume at point $\vect{x}$, the Lorentz force law reads:
\[ \vect{f} = \rho\vect{E} + \vect{J} \times \vect{B} .\]

\subsection{Maxwell's Equations}

Maxwell’s equations are the basis for understanding of all electromagnetic phenomena. They describe space- and time-dependent electric and magnetic fields, and how they are generated by charges and currents. Maxwell’s equations also describe light and other forms of electromagnetic radiation. They are commonly expressed in a differential form or in an integral form. 

\subsubsection{Differential form}
Electric charges whose density is $\rho$ are the sources of the electric field $\vect{E}$. This is expressed by the \emph{Gauss law}: 
\[ 1) \qquad
\divvec{\electricfield}  =  \frac{1}{\epsilon_0} \electricchargedensity .
\]

Field lines of $\vect{B}$ are closed, so the net magnetic charge inside a closed surface $S$ is always zero. This is equivalent to the statement that there are \emph{no magnetic monopoles}. Instead, there are only magnetic dipoles, made of a positive magnetic charge and a negative magnetic charge tied together so they can never be separated. Mathematically this is expressed by the equation:
\[ 2) \qquad \divvec{\magneticfield}  =  0 .\]

The electromotive force around a closed circuit is proportional to the rate of change of flux of the field $\vect{B}$ through the circuit (\emph{Faraday's law}). Faraday's law describes time-varying magnetic fields, and how they produce electric fields. In differential form this law is expressed by the following formula: 
\[
3) \qquad \curlvec{\electricfield} = -\timediff{\magneticfield} .
\]

Electric currents with density $\vect{J}$ are the sources of the magnetic induction field $\vect{B}$. This is expressed by \emph{the Ampère's law}, which describes how time-varying electric fields produce magnetic fields: 
\[  
4) \qquad \curlvec{\magneticfield}  =  \mu_0 \vect{J} + \mu_0\epsilon_0  \timediff{\electricfield} .
\]

These equations involve two constants $\epsilon_0$ and $\mu_0$ that are not themselves of much physical significance, their values appropriate to SI units are fixed by experiment.  

The first is the electric constant:
\[ 
\epsilon_0 \approx 8.85 \times 10^{-12} \mathrm{m^3 \, kg^{-1} \, s^2 \, C} .
\]
It can be thought of as characterising the strength of the electric interactions. The other is the magnetic constant:
\[ 
\mu_0 = 4\pi \times 10^{-7} \mathrm{m \, kg \, C^{-2}} \approx 1.25 \times 10^{-6} \mathrm{m \, kg \, C^{-2}} .
\]

\subsubsection{Integral form}

For better understanding of Maxwell equations, we rewrite them in integral form by using Stokes' theorem:
\[
\oint_C \vect{F} = \int_S \nabla \times \vect{F} \cdot d\vect{S} .
\]

If we rewrite the first Maxwell equation, the Gauss law, in integral form by integrating over fixed volumes using the divergence theorem  or fixed surfaces using Stokes' theorem we get: 
\[ 
\divvec{\electricfield}  =  \frac{1}{\epsilon_0} \electricchargedensity  \Rightarrow  \int_{V} d^{3}x \divvec{\electricfield} = \int_{S} \vect{E} \cdot d\vect{S} = \frac{1}{\epsilon_0} \int_{V} d^{3}x\rho   .
\]

The integral of the charge density over $V$ is simply the total charge $Q = \int_{V} d^{3}x\rho $ contained in the region. The integral of the electric field over $S$ is called the flux through $S$.
\[ 1) \qquad
\int_{S = \partial{V}} \vect{E} \cdot d\vect{S} = \frac{Q}{\epsilon_0} 
\]

The left-hand side is the flux of $\vect{E}$ out of volume $V$. It does not matter what shape the surface $\vect{S}$ takes. As long as it surrounds a total charge $Q$, the flux through the surface will always be $ \frac{Q}{\epsilon_0} $.

Similarly, from $\divvec{\magneticfield}  =  0$ implies that
\[ 2) \qquad
\int_{\partial{V}} \vect{B} \cdot d\vect{S} = 0 
\]
for any closed surface $S = \partial{V}$. This can be interpreted as the statement that there are no magnetic ‘charges’ or magnetic monopoles.

\begin{figure}[!b]
   \centering
  \includegraphics{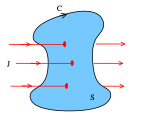}
   \caption{Flux through surface $S$ within the contour $C$.}
   \label{amper_law}
\end{figure}

Next, $\curlvec{\electricfield}  =  - \timediff{\magneticfield}$ implies
\[  3) \qquad
\int_{C}\vect{E} \cdot d\vect{r} = \int_{S} \curlvec{\electricfield} \cdot d\vect{S}  = - \int_{S} \timediff{\magneticfield} \cdot d\vect{S} = -\frac{d}{dt} \int_{S} \vect{B} \cdot d\vect{S} 
\]
by applying Stokes' theorem to a fixed curve $C = \partial{S}$ bounding a fixed open surface $S$ (see Fig.~\ref{amper_law}). If we define the electromotive force (or emf) acting in $C$ by
\[ \varepsilon = \int_{C}\vect{E} \cdot d\vect{r} \]
and the flux of $\vect{B}$ through (the open surface) $S$ by
\[ \Phi =  \int_{S} \vect{B} \cdot d\vect{S} \]
then we get Faraday's Law of induction 
\[ \varepsilon = - \frac{d\Phi}{dt}  .\] 

The electromotive force is the tangential component of the force per unit charge, integrated along the wire. Another way to think about it is as the work done on a unit charge moving around the curve C. If there is a non-zero emf present then the charges will be accelerated around the wire, giving rise to a current.

Finally, the same is repeated for the Ampère's law: 
\[ 4) \qquad
\int_{S}\curlvec{\magneticfield}\cdot d\vect{S}  =  \mu_0 \int_{S}\vect{J} \cdot d\vect{S}  + \mu_0\epsilon_0  \int_{S}\timediff{\electricfield} \cdot d\vect{S} .
\]

Hence, in the case of steady current (no time dependence), Stokes' theorem implies
\[
\int_{C}\vect{B} \cdot d\vect{r} = \mu_0 \int_{S}\vect{J} \cdot d\vect{S} = \mu_0I 
\]
where $I = \int_{S}\vect{J} \cdot d\vect{S}$ is the flux of $J$ through an open surface $S$ bounded by $C$ or the total current through $S$ (or $C$) (see Fig.~\ref{amper_law}).

\subsection{Discontinuity formulas}

With the help of Maxwell's equations, boundary conditions of electromagnetic fields at the interface between different materials can be derived. For a surface $S$ with unit normal $\vect{n}$ which separates regions $V_{\pm}$ of space, with $\vect{n}$ pointing from $S$ into $V_{+}$

\begin{itemize}
    \item if it carries the charge density $\sigma$ per unit area, then 
\[
\vect{n} \cdot \vect{E}|^{+}_{-} = \frac{\sigma}{\epsilon_0}
\]
\[
\vect{n} \wedge \vect{E}|^{+}_{-} = 0
\]
where $\vect{E}|^{+}_{-}$ is the difference of the electric fields on both sides of $S$. $\vect{n}\cdot\vect{v}$ and $\vect{n}\wedge\vect{v}$ give the normal and tangential components of any vector $\vect{v}$, and the tangential component satisfies $\vect{n} \cdot \vect{n} \wedge \vect{v} = 0$.

\item if it carries the current density $\vect{s}$ per unit length (charge crossing unit length in $S$ in unit
time). 
\[
\vect{n} \cdot \vect{B}|^{+}_{-} = 0
\]
\[
\vect{n} \wedge \vect{B}|^{+}_{-} = \mu_0\vect{s}
\]
where $\vect{B}_{\pm}$ is the magnetic fields inside
the $V_{\pm}$ sides of $S$.

\end{itemize}

If we replace $\nabla$ by $\vect{n}|^{+}_{-}$ in Maxwell's equations, then we expect that $\vect{n} \cdot \vect{J}|^{+}_{-} = 0$ at a surface of discontinuity, one that may carry surface density of charge.

These formulas can be applied to deriving
\[
\vect{n} \cdot \vect{B}|^{+}_{-} = 0
\]
\[
\vect{n} \wedge \vect{E}|^{+}_{-} = 0.
\]

\section{Electrostatics}

Now we look at the condition where there are no currents (electrostatics). We assume that the charges are pinned in place and there are forces between charges.  

Since nothing moves, we are looking for time independent solutions to Maxwell’s equations with $\vect{J} = 0$. This means that we can consistently set $\vec{B} = 0$ and we are left with two of Maxwell’s equations:
\[ \divvec{\electricfield}  =  0\]
\[ \curlvec{\electricfield} =  0.\]

\subsection{Coulomb Force}

It can be shown that Gauss law reproduces the more familiar the Coulomb force law. For a particle of some radius $R$ with a spherically symmetric charge
distribution centered at the origin (see Fig. ~\ref{coulomb_force}), the electric field $\vect{E}(x) = E(r)\hat{\vect{r}}$ ($\hat{\vect{r}}$ – radial unit vector) at some radius $r > R$ will be calculated the following way:
\begin{figure}[!bp]
   \centering
  \includegraphics[width=0.3\textwidth]{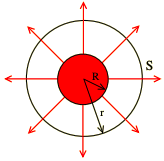}
  \caption{Field around a spherically symmetric charge distribution.}
    \label{coulomb_force}
\end{figure}
\[ 
\int_S \vect{E} \cdot d\vect{S} =  
E(r) \int_S \hat{\vect{r}} \cdot d\vect{S} = 
E(r) 4 \pi r^2= 
\frac{Q}{\epsilon_0} 
\]
where the factor of $4\pi r^2$ has arisen simply because it is the area of the Gaussian sphere. So the electric field outside a spherically symmetric distribution of charge $Q$ is
\[ 
\vect{E}(x) = \frac{Q}{4\pi\epsilon_0r^2} \hat{\vect{r}}. 
\]

From the Lorentz force law we know that a test charge $q$ moving in the region $r>R$ experiences a force
\[ 
\vect{F}(x) = \frac{Q q}{4\pi\epsilon_0r^2} \hat{\vect{r}} .
\]

This is, as it is well known, the Coulomb force between two static charged particles. Notice that $1/\epsilon_0$ characterises the strength of the force. If the two charges have the same sign, so that $Qq > 0$, the force is repulsive, pushing the test charge away from the origin. If the charges have opposite signs, $Qq < 0$, the force is attractive, pointing towards the origin. We see that Gauss law reproduces the Coulomb force.

\subsection{Electrostatic Potential}

From the equation $\curlvec{\electricfield} = 0$ we can see that the electric field can be written as the gradient of some function:
\[ \vect{E} = -\nabla\phi .\]

The scalar $\phi$ is called the \emph{electrostatic potential} or \emph{scalar potential}. If we revert to the original differential form of Gauss law, this takes the form of the Poisson equation:
\[ 
\divvec{\electricfield}  =  
\frac{1}{\epsilon_0} \electricchargedensity \Rightarrow \nabla^2 \phi\ =  
- \frac{1}{\epsilon_0} \electricchargedensity .
\]

In regions of space, where the charge density vanishes, we are left solving the Laplace equation:
\[ \nabla^2 \phi\ =  0 \]
where the operator $\nabla^2$ is called Laplacian:
\[ 
\nabla \cdot \nabla = 
\nabla^2 = 
\frac{\partial^2}{\partial x^2} + \frac{\partial^2}{\partial y^2} + \frac{\partial^2}{\partial z^2} .
\]

Solutions to the Laplace equation are said to be harmonic functions.

\subsection{Principle of superposition}
The Poisson equation is linear in both  $\phi$ and $\rho$. This means that if we know the potential $\phi_1$ for some charge distribution $\rho_1$ and the potential $\phi_2$ for another charge distribution $\rho_2$, then the potential for $\rho_1 + \rho_2$ is simply $\phi_1 + \phi_2$. It means that the electric field for a bunch of charges is just the sum of the fields generated by each charge. This is called the \emph{principle of superposition} for charges.

Similarly, the electric field is just the sum of the electric fields made by the two point charges. This follows from the linearity of the equations and is a simple application of the principle of superposition.

\subsection{Continuous distribution of charges}

Derivation of the potential due to a point charge together with the principle of superposition, is actually enough to solve the potential due to any charge distribution. This is because the solution for a point charge is nothing other than the Green's function for the Laplacian. The Green's function is defined to be the solution to the equation
\[ 
\nabla^2 G(\vect{r};\vect{r'}) = 
\delta^3 (\vect{r} - \vect{r'}  )  
\]
which is 
\[ 
G(\vect{r};\vect{r'}) = 
- \frac{1}{4\pi} \frac{1}{|\vect{r} - \vect{r'}|} .
\] 

We can apply the usual Green's function methods to the general Poisson equation, $\nabla^2 \phi\ =  - \frac{1}{\epsilon_0} \electricchargedensity$. In what follows, well take $\phi(r) \ne 0$ only in some compact region, $V$, of space.

The solution to the Poisson equation is given by
\[ 
\phi(r) = -\frac{1}{\epsilon_0} \int_V d^3\vect{r'}G(\vect{r};\vect{r'}) \rho(\vect{r'}) = -\frac{1}{\epsilon_0} \int_V d^3\vect{r'} \frac{\rho(\vect{r'})}{|\vect{r} - \vect{r'}|} .
\]

Similarly, the electric field arising from a general charge distribution is
\[  
\vect{E}(\vect{r}) = - \nabla \phi(\vect{r}) = \frac{1}{4 \pi \epsilon_0} \int_V d^3\vect{r'} \rho(\vect{r'}) \frac{\vect{r} - \vect{r'}}{|\vect{r} - \vect{r'}|^3} .
\]

\section{Magnetostatics}

Charges give rise to electric fields. Current give rise to magnetic fields. The magnetic fields is induced by steady currents. In this case the charge density $\rho = 0$, so $E = 0$. The time independent solutions to the Maxwell equations in case of magnetic fields will be:
\[  \curlvec{\magneticfield}  =  \mu_0 \vect{J}\]
\[ \divvec{\magneticfield}  =  0 .\]
If the current density $\vect{J}$ is fixed, these equations have a unique solution. 

\subsection{Steady Currents}
Because $\rho = 0$, there cannot be any net charge, but we need moving charge for magnetic fields. This means that we necessarily have both positive and negative charges which balance out at all points in space. Nonetheless, these charges can move so there is a current even though there is no net charge transport.

It is exactly what happens in a typical wire. In that case, there is background of positive charge due to the lattice of ions in the metal. Meanwhile, the electrons are free to move, but they all move together, so that at each point we still have $\rho = 0$. The continuity equation, which captures the conservation of electric charge, is
\[ 
\timediff{\rho} + \nabla \cdot \vect{J}= 0 .
\]
Since the charge density is unchanging (and indeed, vanishing), we have $\nabla \cdot \vect{J}= 0$.

Mathematically, this is just saying that if a current flows into some region of space, an equal current must flow out to avoid the build up of charge. This is consistent with $\curlvec{\magneticfield}  =  \mu_0 \vect{J}$  since, for any vector field, $ \nabla \cdot ( \nabla \times \vect{B}) = 0$. 

\subsection{Ampère's Law}

The first equation of magnetostatics,
\[
\curlvec{\magneticfield}  =  \mu_0 \vect{J}
\]
is known as Ampère's law. An equivalent integral form  over some open surface $S$ with boundary $C = \partial S$ (see Fig.~\ref{amper_law}). By using Stokes' theorem: 
\[
\int_S \nabla \times \vect{B} \cdot d\vect{S} =  \oint_C \vec{B} \cdot d\vect{r} = \mu_0 \int_S \vect{J} \cdot d\vect{S} .
\]

The surface $S$ comes with a normal vector $\hat{\vect{n}}$ which points away from $S$ in one direction. The line integral around the boundary is done in the right-handed sense, meaning that if you stick the thumb of your right hand in the direction $\hat{\vect{n}}$ then your fingers curl in the direction of the line integral.

The integral of the current density over the surface $S$ is the same thing as the total current $I$ that passes through $S$. Ampère's law in integral form then reads: 
\[
 \oint_C \vec{B} \cdot d\vect{r} = \mu_0 I .
\]
\subsection{Vector Potential}

We are guaranteed a solution to $\nabla \cdot \vect{B} = 0$ if we write the magnetic field as the curl of some vector field:
\[
\vect{B} = \nabla \times \vect{A}
\]
where $\vect{A}$ is called the vector potential. 

Then Ampère's law becomes:
\[
\nabla \times \vect{B} = -\nabla^2\vect{A} + \nabla(\nabla \cdot \vect{A}) = \mu_0\vect{J} .
\]

The choice of $\vect{A}$ above is not unique as there are lots of different vector potentials and they give rise to the same magnetic field $\vect{B}$. This is because the curl of a gradient is automatically zero. This means that we can always add any vector potential of the form  $\nabla \chi$ for some function $\chi$ and the magnetic field remains the same:
\[
\vect{A'} = \vect{A} + \nabla \chi \rightarrow  \nabla  \times \vect{A'} =  \nabla  \times \vect{A} .
\]

We can always find a gauge transformation $\chi$ such that  $\vect{A'}$ satisfies $\nabla \cdot \vect{A'} = 0$. This choice is usually referred to as \emph{Coulomb gauge}. 

From now, we will always assume that we are working in Coulomb gauge and our vector potential obeys $\nabla \cdot \vect{A} = 0$.

\subsection{Biot-Savart law}

From the Ampère law by using the vector potential and Coulomb gauge we solve the equation for the magnetic field $B$ in the presence of a general current distribution:
\[
-\nabla^2\vect{A} = \mu_0\vect{J} .
\]
The most general solution using Green's functions: 
\[
\vect{A}(\vect{x}) = \frac{\mu_0}{4\pi}\int_V d^3x'\frac{\vect{J(\vect{x})}}{|\vect{x - \vect{x'}}|} .
\]

This gives a practical solution for the magnetic field $\vect{B} = \nabla \times \vect{A}$. We need to remember that the $\nabla$ acts on the $\vect{x}$ rather than the $\vect{x'}$. We find:
\[
\vect{B}(\vect{x}) = \frac{\mu_0}{4\pi}\int_V d^3x'\frac{\vect{J(\vect{x})} \times (\vect{x}- \vect{x'})}{|\vect{x - \vect{x'}}|^3} .
\]

This is known as the \emph{Biot-Savart law}. It describes the magnetic field due to a general current density.

\subsection {Motion of a charged particle}

As known, the position $\vect{r}(t)$ of a particle of charge $q$ is dictated by the electric and magnetic fields through the Lorentz force law
\[ \vect{F} = q (\vect{E} + \vect{v} \times \vect{B}) .\]

\begin{figure}[htbp]
   \centering
  \includegraphics[width=0.35\textwidth]{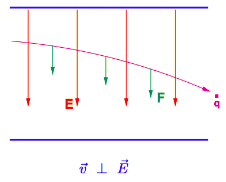}
    \includegraphics[width=0.25\textwidth]{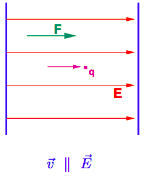}
      \includegraphics[width=0.45\textwidth]{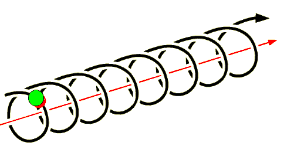}
  \caption{Motion of a charged particle in electric (above) and magnetic (below) fields. In case of an electric  field, the force is always in the direction of the field, also for particles in rest. For a magnetic field, the force is perpendicular to both $\vect{v}$ and $\vect{B}$.}
\label{particle_motion}
\end{figure}

In case of absence of the magnetic fields the second term $\dot{\vect{r}} \times \vect{B}$ vanishes, $\vect{F} = q (\vect{E} + \dot{\vect{r}} \times \xcancel{\vect{B}})$,  i.e. only electric fields are taken into account, and we have
\[ 
\frac{d}{dt}(m\vect{v}) = \vect{F} = q \vect{E} .
\]
The force is acting always in the direction of the electric field, $\vect{E}$, also for particles at rest.

The second case, absence of the electric field, only magnetic field is acting, $\vect{F} = q (\xcancel{\vect{E}} + \vect{v} \times \vect{B})$, we get 
\[ 
\frac{d}{dt}(m\vect{v}) = \vect{F} = q \cdot \vect{v} \times \vect{B} .
\]
In this case the force is perpendicular to both $\vect{v}$ and $\vect{B}$ (see Fig.~\ref{particle_motion}).

\section{Electromagnetism}

For static situations, Maxwell’s equations split into the equations of electrostatics and  and the equations of magnetostatics. The only hint that there is a relationship between electric and magnetic fields comes from the fact that they are both sourced by charge: electric fields by stationary charge; magnetic fields by moving
charge. However, the connection becomes more direct when things change with time.

\subsection{Faraday's Law of Induction}

One of the Maxwell equations relates time varying magnetic fields to electric fields: 
\[
\curlvec{E} + \timediff{\vect{B}} = 0 .
\]

If we change a magnetic field, we will create an electric field. In turn, this electric field can be used to accelerate charges which create a current. The process of creating a current through changing magnetic fields is called \emph{induction}. 

\emph{Faraday's law} describes time-varying magnetic fields, and how they produce electric fields. In integral form this law is expressed by the following formula
\[
\int_C \vect{E} d\vect{r }=  -\frac{d}{dt}\int_S \vect{B} \cdot d\vect{S} .
\]

Faraday's law of induction can be expressed through the electromotive force, $\varepsilon = \int_{C}\vect{E} \cdot d\vect{r}$ and flux of  $\vect{B}$ through (the open surface) $S$, $\Phi =  \int_{S} \vect{B} \cdot d\vect{S}$:
\[ \varepsilon = - \frac{d\Phi}{dt}  .\] 

\subsection{Wave Equation}

Maxwell’s equations have wave-like solutions for the electric and magnetic fields in free space. The electric field must solve the wave equation
\[
\nabla^2 \vect{E} - \frac{1}{c^2}\frac{\partial^2 \vect{E}}{\partial t^2} = 0 .
\]

For the vacuum or free-space, i.e. $\rho = 0$, $\vect{J} = 0$, the solution of wave equations can be written in the form:
\[
\vect{E} = \vect{E_0} \sin(\omega t - \vect{k}\vect{r}) 
\]
where 
\begin{itemize}
    \item $\vect{k}$ is the wave vector with $|\vect{k}| = k$, which gives the direction of propagation of the wave.
    \item The quantity $\omega$ is called the angular frequency and is taken to be positive. The actual frequency $f = \frac{\omega}{2\pi}$ measures how often a wave peak passes by. $\omega$ relates to $k$ by $\omega^2 = c^2k^2$.
    \item The period of oscillation is $T = \frac{2\pi}{\omega}$.
    \item The wavelength of the wave is $\lambda = \frac{2\pi}{k}$.
     The wavelength of visible light is between
     $\lambda \approx 3.9 \times 10^{-7} \mathrm{m}$ and $7 \times  10^{-7} \mathrm{m} $ . At one end of the spectrum, gamma rays have wavelength  $\lambda \approx 10^{-12} \mathrm{m}$ and X-rays around $\lambda \approx 10^{-10}$ to $10^{-8} \mathrm{m}$. At the other end, radio waves have $\lambda \approx 1\:\mathrm{cm}$ to $10\:\mathrm{km}$. Of course, the electromagnetic spectrum does not stop at these two ends. Solutions exist for all $\lambda$.
    \item $\vect{E_0}$ is the amplitude of the wave.
    \item The phase velocity $c$ of the wave is given by the \emph{dispersion} relation:
    \[
    c = \frac{\omega}{|k|} = \frac{1}{\sqrt{\mu_0 \epsilon_0}} 
    \]
   
\end{itemize}

The linearity of the Maxwell equations allows to write the solutions
in complex notation:
\[
\vect{E}(\vect{r, t}) = \vect{E_0} e^{i(\omega t - \vect{k}\vect{r}) } 
\]
\[
\vect{B}(\vect{r, t}) = \vect{B_0} e^{i(\omega t - \vect{k}\vect{r})}
\]
 where $\vect{E_0}$, $\vect{B_0}$ are  the amplitudes of the electric and magnetic fields, respectively, $\vect{k}$ their wave vector and $\omega$ its angular frequency.

\begin{figure}[htbp]
\centering
  \includegraphics[width=0.5\textwidth]{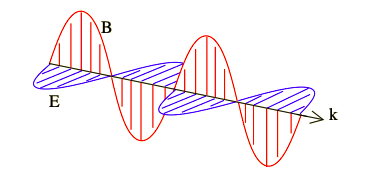}
  \caption{The electromagnetic wave.}
\label{em_wave}
\end{figure}

$\nabla \cdot \vect{E} = 0$ implies $\vect{E_0} \cdot \vect{k} = 0$, and likewise $\nabla \cdot \vect{B} = 0$ 
implies $\vect{B_0} \cdot \vect{k} = 0$, so that both these fields, magnetic and electric, are transverse to the direction of propagation, $\vect{B} \perp \vect{E} \perp \vect{k}$ (see Fig~\ref{em_wave}).

\subsection{Propagation of electromagnetic waves in a conductor}



In a conducting medium of conductivity $\sigma$  we have 
\[ 
\vect{J}= \sigma \vect{E}
\]
and hence
\[ 
\nabla \cdot \vect{J} = \sigma \nabla \cdot \vect{E} = \frac{\sigma}{\epsilon_0} \rho .
\]

So, we can rewrite the continuity equation,
\[ 
\timediff{\rho} + \nabla \cdot \vect{J}= 0 
\]
as
\[ 
\timediff{\rho} + \frac{\sigma}{\epsilon_0} \rho = 0 
\]
and
\[
\rho (t) = \rho_0 \exp{-\frac{t}{\tau}} .
\]

The ratio
\[
\tau = \frac{\epsilon_0}{\sigma}
\]
is called the \emph{relaxation time} of the conducting medium. For perfect conductors, $\sigma = \infty$, so that the relaxation time is vanishing. 

For good, but not perfect, conductors $\frac{\sigma}{\epsilon} \approx 10^{14} \mathrm{sec}^{-1}$, so and charges move almost instantly to the surface of the conductor. In this case $\tau$ is small of the order of $10^{-14} \mathrm{sec}$. For times much larger than the relaxation time there are practically no charges inside the conductor. All of them have moved to its surface where they form a charge density. 

For the case of an isolator $\sigma = 0$. In this case the solution of the wave equation is reduces to an ordinary plane wave which is propagating with wave vector $\vect{k}$. 

For a very good conductor, which includes the case of a perfect conductor, the conductivity is large, so that the relaxation time is vanishing. 

Therefore inside a good conductor the field is attenuated in the direction of the propagation and its magnitude decreases exponentially as it penetrates into the conductor. $\delta$ is a constant called the \emph{skin depth}, given by the following expression: 
\[
\delta = \sqrt{\frac{2}{\mu_r\sigma \omega}} .
\]

The depth of the penetration is set by $\delta$ and  is smaller the higher the conductivity, the higher the permeability and the frequency. 
Skin depths for a good conductor (metal) is $\delta \approx 10\:\mathrm{\mu  m}$ at $\omega = 50\:\mathrm{MHz}$.

\subsection{Propagation of electromagnetic waves in a  highly conductive materials}

\begin{figure}[b]
\centering
  \includegraphics[width=0.8\textwidth]{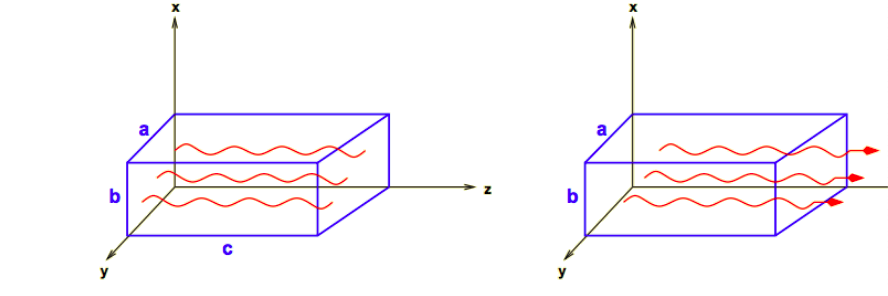}
  \caption{Rectangular, conducting cavities (left) and wave guides (right) of dimensions $a \times b\times c$ and $a \times b$}
\label{rf_wg}
\end{figure}

It's very important to understand electromagnetic waves in a highly conductive material (see Fig.~\ref{rf_wg}) to be able to design RF cavities and wave-guides  for charged-particle accelerators. 

The fields $\vect{E}$ and $\vect{B}$ are zero inside perfectly conducting media, it therefore follows the boundary conditions at $z = 0$ are: 
\[ \vect{E}_{||} = 0, \vect{B}_\perp = 0 .\]

\pagebreak
This implies that
\begin{itemize}
    \item all energy of an electromagnetic wave is reflected from the surface of an ideal conductor;
    \item  fields at any point in the ideal conductor are zero;
    \item only some field patterns are allowed in wave guides and RF cavities (examples of highly superconductive material in accelerator physics).
\end{itemize}

\subsubsection{RF Cavities}
The  word  ‘cavity’  is  derived  from  the  Latin  \emph{cavus} (hollow) –  a  void  or  empty  space  within  a  solid  body. The  purpose  of  an  RF  cavity  is  to  interact  with  charged  particle  beams  in  an  accelerator.  For  the particles  to  gain  energy,  this  interaction  is  the  acceleration  in  the  direction  of  particle  motion,  but   special cavities exist to bunch, or de-bunch the beam, others to decelerate particles or to kick  them sideways. For all these different applications, optimization of  the  cavity  for  the  given  application has to be done. 

Maxwell’s equations, along with the boundary conditions of an empty cavity,  have solutions with non-vanishing fields even if no sources are present:

\[
E_x = E_{x0} \cdot \cos(k_xx)  \cdot \sin(k_y y) \cdot \sin(k_z z) \cdot e^{i\omega t}
\]
\[
E_y = E_{y0} \cdot \sin(k_xx)  \cdot \cos(k_y y) \cdot \sin(k_z z) \cdot e^{i\omega t}
\]
\[
E_z = E_{z0} \cdot \sin(k_xx)  \cdot \sin(k_y y) \cdot \cos(k_z z) \cdot e^{i\omega t}
\]

\[
B_x = \frac{i}{\omega} (E_{y0}k_z - E_{z0}k_y) \cdot sin(k_xx)  \cdot \cos(k_y y) \cdot cos(k_z z) \cdot e^{i\omega t}
\]
\[
B_y = \frac{i}{\omega} (E_{z0}k_x - E_{x0}k_z) \cdot \cos(k_xx)  \cdot \sin(k_y y) \cdot \cos(k_z z) \cdot e^{i\omega t}
\]
\[
B_z = \frac{i}{\omega} (E_{x0}k_y - E_{y0}k_x) \cdot \cos(k_xx)  \cdot \cos(k_y y) \cdot \sin(k_z z) \cdot e^{i\omega t} .
\]
Notice, that there is the cosine dependence on the coordinate corresponding to the component of the field  and the sine dependence on the other coordinates in the electric field.

To satisfy the wave equations the wave vector and frequency must be related: 
\[
k_x^2 + k_y^2 + k_z^2  = \frac{\omega^2}{c^2} .
\]


At the boundaries of the conducting cavities field must be zero, i.e. the tangential component of the electric field vanishes at the conducting wall. This is only possible with the following constrains on the wave vector: 
\[
k_x = \frac{m_x\pi}{a}; \quad k_y = \frac{m_y\pi}{b};  \quad k_z = \frac{m_z\pi}{c}, 
\]
where $m_{x,y,z}$ – are integer numbers, which are called mode numbers. They specify the dependence of the electric field on the coordinate and they are very important values in design of RF cavities. Note that at least two mode numbers should be non-zero. Otherwise the field vanishes everywhere. 

Since the components of the wave vector are constrained to discrete values (since the mode numbers must be integer), the frequency is only allowed to take the certain values: 
\[
\omega = \pi c \sqrt{\frac{m_x^2}{a^2} + \frac{m_y^2}{b^2} +  \frac{m_z^2}{c^2}} .
\]

\subsubsection{Wave Guides}
A wave guide is a metallic open ended tube of arbitrary cross sectional shape. Under certain conditions electromagnetic waves can propagate along its axis. It's a perfectly conducting tube with rectangular cross-section of height and width $a$ and $b$. This is essentially a cavity resonator with length $c \rightarrow \infty$. The tube can be filled with a nondissipative medium characterized by dielectric constant $\epsilon$  and magnetic permeability $\mu$. 

Together with Maxwell's equations the electric field must solve the wave equation
\[
\nabla^2 \vect{E} - \frac{1}{c^2}\frac{\partial^2 \vect{E}}{\partial t^2} = 0 .
\]

The solution will be in a form: 

\[
E_x = E_{x0} \cdot \cos(k_xx) \cdot \sin(k_y y)  \cdot e^{i(\omega t - k_zz)}
\]
\[
E_y = E_{y0} \cdot \sin(k_xx) \cdot \cos(k_y y)  \cdot e^{i(\omega t - k_zz)}
\]
\[
E_z = -iE_{z0} \cdot \sin(k_xx) \cdot \sin(k_y y) \cdot e^{i(\omega t - k_zz)} .
\]

By  applying the boundary condition (vanishing components $E_x$, $E_y$, where they are tangential to the wall) we find that $k_x$ and $k_y$ for any integer mode number $m$ must be:
\[
k_x = \frac{m_x\pi}{a}; \quad k_y = \frac{m_y\pi}{b}.
\]

The longitudinal component, $E_z$, always vanishes on the walls, so there is no constraint on $k_z$. 

Again, we have to satisfy Maxwell's equation $\nabla \cdot \vect{E} = 0$. This leads to a relation between the amplitudes and the components of the wave vector:
\[
k_xE_{x0} + k_yE_{y0} +k_zE_{z0} = 0.
\]

Together with the wave equation it leads to the dispersion relation:
\[
k_x^2 + k_y^2 + k_z^2 = \frac{\omega}{c^2} ,
\]
where $c = \frac{1}{\sqrt{\mu\epsilon}}$.

As there is no constraint on $k_z$ in a waveguide, there is a continuous range of frequencies allowed in a waveguide. However, there is still a \emph{minimum} frequency allowed in a given mode.

$k_z$ must be real for a travelling wave, i.e. $k_z^2 \geq 0$.  The minimum frequency for a propagation wave is called the \emph{cut-off frequency}, $\omega_{\mathrm{CO}}$:
\[
\omega_\mathrm{CO} = \pi c \sqrt{\frac{m_x^2}{a^2} + \frac{m_y^2}{b^2}}.
\]

It is possible for fields to oscillate in a waveguide at a frequency below cut-off frequency. However, such field does not constitute travelling waves. 




Usually waveguides are used so either the electric or magnetic field has no longitudinal component. We shall distinguish the following special modes of propagation: 

\begin{itemize}
    \item Transverse Electric (TE), in which there is no longitudinal component, $E_z$ , of the electric field. Besides having $E_z = 0$ , the appropriate boundary conditions on the walls of the guide dictate that the directional derivative of the z-components of the magnetic field on the conducting wall vanishes. Thus for the TE modes we have:
    \[
    E_z = 0 \quad \mathrm{everywhere}; \quad  B_z \ne 0 ,
    \]
    
    \item Transverse Magnetic (TM), in which case there is no longitudinal component of the magnetic field. In this case we have: 
    \[
    B_z = 0 \quad \mathrm{everywhere}; \quad  E_z \ne 0 ,
    \]
   \item Transverse ElectroMagnetic (TEM) in which both electric and magnetic components are transverse to the wave guide axis. Thus: 
     \[
     E_z, B_z = 0 \quad \mathrm{everywhere} .
    \]
    
\end{itemize}


It can be proven that a hollow wave guide, whose walls are perfect conductors, can not support propagation of TEM waves.

\end{document}